\errorstopmode
\input amssym.def
\input amssym.tex


\magnification=\magstephalf
\hsize=14.0 true cm
\vsize=19 true cm
\hoffset=1.0 true cm
\voffset=2.0 true cm

\abovedisplayskip=12pt plus 3pt minus 3pt
\belowdisplayskip=12pt plus 3pt minus 3pt
\parindent=1.0em


\font\sixrm=cmr6
\font\eightrm=cmr8
\font\ninerm=cmr9

\font\sixi=cmmi6
\font\eighti=cmmi8
\font\ninei=cmmi9

\font\sixsy=cmsy6
\font\eightsy=cmsy8
\font\ninesy=cmsy9

\font\sixbf=cmbx6
\font\eightbf=cmbx8
\font\ninebf=cmbx9

\font\eightit=cmti8
\font\nineit=cmti9

\font\eightsl=cmsl8
\font\ninesl=cmsl9

\font\sixss=cmss8 at 8 true pt
\font\sevenss=cmss9 at 9 true pt
\font\eightss=cmss8
\font\niness=cmss9
\font\tenss=cmss10

\font\sixmib=cmmib6
\font\sevenmib=cmmib7
\font\eightmib=cmmib8
\font\ninemib=cmmib9
\font\tenmib=cmmib10

 at 12 true pt
 at 12 true pt
\font\bigrm=cmr10 at 12 true pt
 at 12 true pt
 at 12 true pt

 at 16 true pt
 at 16 true pt
\font\Bigrm=cmr12 at 16 true pt
 at 16 true pt
 at 16 true pt

\catcode`@=11
\newfam\ssfam
\newfam\mibfam

\def\tenpoint{\def\rm{\fam0\tenrm}%
    \textfont0=\tenrm \scriptfont0=\sevenrm \scriptscriptfont0=\fiverm
    \textfont1=\teni  \scriptfont1=\seveni  \scriptscriptfont1=\fivei
    \textfont2=\tensy \scriptfont2=\sevensy \scriptscriptfont2=\fivesy
    \textfont3=\tenex \scriptfont3=\tenex   \scriptscriptfont3=\tenex
    \textfont\itfam=\tenit                  \def\it{\fam\itfam\tenit}%
    \textfont\slfam=\tensl                  \def\sl{\fam\slfam\tensl}%
    \textfont\bffam=\tenbf \scriptfont\bffam=\sevenbf
                           \scriptscriptfont\bffam=\fivebf
                           \def\bf{\fam\bffam\tenbf}%
    \textfont\ssfam=\tenss \scriptfont\ssfam=\sevenss
                           \scriptscriptfont\ssfam=\sevenss
                           \def\ss{\fam\ssfam\tenss}%
    \textfont\mibfam=\tenmib \scriptfont\mibfam=\sevenmib
                             \scriptscriptfont\mibfam=\sevenmib
                             \def\mib{\fam\mibfam\tenmib}%
    \normalbaselineskip=13pt
    \setbox\strutbox=\hbox{\vrule height8.5pt depth3.5pt width0pt}%
    \let\big=\tenbig
    \normalbaselines\rm}

\def\ninepoint{\def\rm{\fam0\ninerm}%
    \textfont0=\ninerm      \scriptfont0=\sixrm
                            \scriptscriptfont0=\fiverm
    \textfont1=\ninei       \scriptfont1=\sixi
                            \scriptscriptfont1=\fivei
    \textfont2=\ninesy      \scriptfont2=\sixsy
                            \scriptscriptfont2=\fivesy
    \textfont3=\tenex       \scriptfont3=\tenex
                            \scriptscriptfont3=\tenex
    \textfont\itfam=\nineit \def\it{\fam\itfam\nineit}%
    \textfont\slfam=\ninesl \def\sl{\fam\slfam\ninesl}%
    \textfont\bffam=\ninebf \scriptfont\bffam=\sixbf
                            \scriptscriptfont\bffam=\fivebf
                            \def\bf{\fam\bffam\ninebf}%
    \textfont\ssfam=\niness \scriptfont\ssfam=\sixss
                            \scriptscriptfont\ssfam=\sixss
                            \def\ss{\fam\ssfam\niness}%
    \textfont\mibfam=\ninemib \scriptfont\mibfam=\sixmib
                            \scriptscriptfont\mibfam=\sixmib
                            \def\mib{\fam\mibfam\ninemib}%
    \normalbaselineskip=12pt
    \setbox\strutbox=\hbox{\vrule height8.0pt depth3.0pt width0pt}%
    \let\big=\ninebig
    \normalbaselines\rm}

\def\eightpoint{\def\rm{\fam0\eightrm}%
    \textfont0=\eightrm      \scriptfont0=\sixrm
                             \scriptscriptfont0=\fiverm
    \textfont1=\eighti       \scriptfont1=\sixi
                             \scriptscriptfont1=\fivei
    \textfont2=\eightsy      \scriptfont2=\sixsy
                             \scriptscriptfont2=\fivesy
    \textfont3=\tenex        \scriptfont3=\tenex
                             \scriptscriptfont3=\tenex
    \textfont\itfam=\eightit \def\it{\fam\itfam\eightit}%
    \textfont\slfam=\eightsl \def\sl{\fam\slfam\eightsl}%
    \textfont\bffam=\eightbf \scriptfont\bffam=\sixbf
                             \scriptscriptfont\bffam=\fivebf
                             \def\bf{\fam\bffam\eightbf}%
    \textfont\ssfam=\eightss \scriptfont\ssfam=\sixss
                             \scriptscriptfont\ssfam=\sixss
                             \def\ss{\fam\ssfam\eightss}%
    \textfont\mibfam=\eightmib \scriptfont\mibfam=\sixmib
                             \scriptscriptfont\mibfam=\sixmib
                             \def\mib{\fam\mibfam\eightmib}%
    \normalbaselineskip=10pt
    \setbox\strutbox=\hbox{\vrule height7.0pt depth2.0pt width0pt}%
    \let\big=\eightbig
    \normalbaselines\rm}

\def\tenbig#1{{\hbox{$\left#1\vbox to8.5pt{}\right.\n@space$}}}
\def\ninebig#1{{\hbox{$\textfont0=\tenrm\textfont2=\tensy
                       \left#1\vbox to7.25pt{}\right.\n@space$}}}
\def\eightbig#1{{\hbox{$\textfont0=\ninerm\textfont2=\ninesy
                       \left#1\vbox to6.5pt{}\right.\n@space$}}}

\font\sectionfont=cmbx10
\font\subsectionfont=cmti10

\def\figurecaptionfont{\ninepoint}
\def\tablecaptionfont{\ninepoint}
\def\footnotefont{\eightpoint}


\newcount\equationno
\newcount\bibitemno
\newcount\figureno
\newcount\tableno

\equationno=0
\bibitemno=0
\figureno=0
\tableno=0


\footline={\ifnum\pageno=0{\hfil}\else
{\hss\rm\the\pageno\hss}\fi}


\def\section #1. #2 \par
{\vskip0pt plus .10\vsize\penalty-100 \vskip0pt plus-.10\vsize
\vskip 1.6 true cm plus 0.2 true cm minus 0.2 true cm
\global\def\equationlabel{#1}
\global\equationno=0
\leftline{\sectionfont #1. #2}\par
\immediate\write\terminal{Section #1. #2}
\vskip 0.7 true cm plus 0.1 true cm minus 0.1 true cm
\noindent}


\def\subsection #1 \par
{\vskip0pt plus 0.8 true cm\penalty-50 \vskip0pt plus-0.8 true cm
\vskip2.5ex plus 0.1ex minus 0.1ex
\leftline{\subsectionfont #1}\par
\immediate\write\terminal{Subsection #1}
\vskip1.0ex plus 0.1ex minus 0.1ex
\noindent}


\def\appendix #1. #2 \par
{\vskip0pt plus .20\vsize\penalty-100 \vskip0pt plus-.20\vsize
\vskip 1.6 true cm plus 0.2 true cm minus 0.2 true cm
\global\def\equationlabel{\hbox{\rm#1}}
\global\equationno=0
\leftline{\sectionfont Appendix #1. #2}\par
\immediate\write\terminal{Appendix #1. #2}
\vskip 0.7 true cm plus 0.1 true cm minus 0.1 true cm
\noindent}



\def\equation#1{$$\displaylines{\qquad #1}$$}
\def\enum{\global\advance\equationno by 1
\hfill\llap{{\rm(\equationlabel.\the\equationno)}}}

\def\next#1{\cr\noalign{\vskip#1}\qquad}


\def\ifundefined#1{\expandafter\ifx\csname#1\endcsname\relax}

\def\ref#1{\ifundefined{#1}?\immediate\write\terminal{unknown reference
on page \the\pageno}\else\csname#1\endcsname\fi}

\newwrite\terminal

\def\bibitem#1#2\par{\global\advance\bibitemno by 1
\item{[\the\bibitemno]}#2\par}

\def\beginbibliography{
\vskip0pt plus .15\vsize\penalty-100 \vskip0pt plus-.15\vsize
\vskip 1.2 true cm plus 0.2 true cm minus 0.2 true cm
\leftline{\sectionfont References}\par
\immediate\write\terminal{References}
\frenchspacing\parindent=1.8em
\vskip 0.5 true cm plus 0.1 true cm minus 0.1 true cm}

\def\endbibliography{
\nonfrenchspacing\parindent=1.0em}


\def\figurecaption#1{\global\advance\figureno by 1
\narrower\figurecaptionfont
Fig.~\the\figureno. #1}

\def\tablecaption#1{\global\advance\tableno by 1
\vbox to 0.25 true cm { }
\centerline{\tablecaptionfont%
Table~\the\tableno. #1}
\vskip-0.4 true cm}

\def\thicktablerule{\hrule height1pt}
\def\thintablerule{\hrule height0.4pt}

\tenpoint


\def\thismonth{\ifcase\month\or
January\or February\or March\or April\or May\or June\or
July\or August\or September\or October\or November\or December\fi}

\input epsf
\epsfclipon



\def\rmd{{\rm d}}

\def\rme{{\rm e}}



\def\proof{\noindent{\sl Proof:}\kern0.6em}

\def\frac#1#2{\hbox{$#1\over#2$}}
\def\dual{\mathstrut^*\kern-0.1em}

\def\lvec#1{\setbox0=\hbox{$#1$}
    \setbox1=\hbox{$\scriptstyle\leftarrow$}
    #1\kern-\wd0\smash{
    \raise\ht0\hbox{$\raise1pt\hbox{$\scriptstyle\leftarrow$}$}}
    \kern-\wd1\kern\wd0}
\def\rvec#1{\setbox0=\hbox{$#1$}
    \setbox1=\hbox{$\scriptstyle\rightarrow$}
    #1\kern-\wd0\smash{
    \raise\ht0\hbox{$\raise1pt\hbox{$\scriptstyle\rightarrow$}$}}
    \kern-\wd1\kern\wd0}
\def\slash#1{\setbox0=\hbox{$#1$}\setbox1=\hbox{$\kern1pt/$}
    #1\kern-\wd0\kern1pt/\kern-\wd1\kern\wd0}


\def\nabstar#1{{\nabla\kern0.5pt\smash{\raise 4.5pt\hbox{$\ast$}}
               \kern-5.5pt_{#1}}}

\def\drvstar#1{{\partial\kern0.5pt\smash{\raise 4.5pt\hbox{$\ast$}}
               \kern-6.0pt_{#1}}}

\def\ldrvstar#1{{\lvec{\,\partial}\kern-0.5pt\smash{\raise 4.5pt\hbox{$\ast$}}
               \kern-5.0pt_{#1}}}


\def\MeV{{\rm MeV}}

\def\fm{{\rm fm}}



\def\rbar{\bar{r}}
\def\sbar{\bar{s}}


\def\dirac#1{\gamma_{#1}}
\def\diracstar#1#2{
    \setbox0=\hbox{$\gamma$}\setbox1=\hbox{$\gamma_{#1}$}
    \gamma_{#1}\kern-\wd1\kern\wd0
    \smash{\raise4.5pt\hbox{$\scriptstyle#2$}}}


\def\Ad{{\rm Ad}\kern0.1em}


\def\ms{m_s}
\def\mpi{M_{\pi}}

\def\Fpi{F_{\pi}}
\def\FK{F_K}
\def\FKref{F_{K,{\rm ref}}}
\def\mbar{\overline{m\kern-0.07em}\kern0.07em}
\def\Meff{M_{\rm eff}}
\def\Mpi{\mpi}
\def\Kstar{K^{\ast}}
\def\MK{M_{K}}
\def\MKstar{M_{\Kstar}}
\def\MKref{M_{K,\rm ref}}
\def\mref{m_{\rm ref}}
\def\msref{m_{s,{\rm ref}}}
\def\Fhat{\hat{F}}
\def\Bhat{\hat{B}}
\def\ZA{Z_{\rm A}}


\def\csw{c_{\rm sw}}
\def\cA{c_{\rm A}}
\def\bA{b_{\rm A}}
\def\Ncfg{N_{\rm cfg}}
\def\Nops{N_{\rm op}}
\def\lbar{\bar{l}}
\def\lhat{\hat{l}}

%
\rightline{CERN-PH-TH/2006-198}

\vskip0.8cm 
\centerline{\Bigrm
QCD with light Wilson quarks on fine lattices (I):
}
\vskip 0.3 true cm
\centerline{\Bigrm
first experiences and physics results
}
\vskip 0.6cm
\centerline{\bigrm L.~Del Debbio$^{1}$, L.~Giusti$^{2,}$%
\footnote{$^{\ast}$}{\footnotefont%
On leave from
Centre de Physique Th\'eorique, CNRS Luminy, F-13288 Marseille, France},
M.~L\"uscher$^2$, R.~Petronzio$^3$, N.~Tantalo$^{3,4}$}
\vskip2.0ex
\centerline{$^1\hskip-3pt$ \it
SUPA, School of Physics, University of Edinburgh, Edinburgh EH9 3JZ, UK}
\vskip1.0ex
\centerline{$^2\hskip-3pt$ \it
CERN, Physics Department, TH Division, CH-1211 Geneva 23, Switzerland}
\vskip1.0ex
\centerline{$^3\hskip-3pt$ \it
Universit\`a di Roma ``Tor Vergata'' and INFN sezione ``Tor Vergata'',}
\centerline{\it Via della Ricerca Scientifica 1,  I-00133 Rome, Italy}
\vskip1.0ex
\centerline{$^4\hskip-3pt$ \it
Centro Enrico Fermi, Via Panisperna 89 A, I-00184 Rome, Italy}
\vskip 0.8 true cm
\thintablerule
\vskip 2.0ex
\ninepoint
\leftline{\bf Abstract}
\vskip 1.0ex\noindent
Recent conceptual, algorithmic and technical advances
allow numerical simulations of lattice QCD with Wilson quarks 
to be performed at significantly smaller quark masses than 
was possible before.
Here we report on simulations of two-flavour QCD 
at sea-quark masses from slightly above to approximately $1/4$ of the
strange-quark mass, on lattices with up to $64\times32^3$ points 
and spacings from $0.05$ to $0.08$ fm.
Physical sea-quark effects are clearly seen on these lattices,
while the lattice effects appear
to be quite small, even without O($a$) improvement.
A striking result is that
the dependence of the pion mass on the sea-quark mass is
accurately described by leading-order chiral perturbation theory
up to meson masses of about $500$ MeV. 
\vskip 2.0ex
\thintablerule

\tenpoint

\vskip-0.2cm

\section 1. Introduction

Many different formulations of lattice QCD are currently in use. 
The aim to reduce the lattice effects and to 
preserve chiral symmetry as much as possible
has been the principal motivation for 
the introduction of increasingly complicated lattice actions.
Highly improved actions are not 
obviously the best choice in practice, however, 
since they tend to slow down the 
numerical simulations by a large factor. 
Moreover, the conceptual transparency
of the lattice theory (otherwise one of its greatest assets)
may be compromised in extreme cases.

The philosophy advocated here is to keep the theory simple 
at the fundamental level (thus sticking to Wilson's formulation
[\ref{Wilson}] and its close relatives)
and to develop adapted computational 
strategies that allow simulations of large lattices with 
small spacings to be performed efficiently. Extrapolations
to the continuum limit will then still be required, but the 
hope is that, in many cases of interest,
the lattice effects will already be small
at the accessible lattice spacings.

For many years, the Wilson theory
had the reputation of being difficult to simulate at light-quark 
masses significantly smaller than half the strange-quark mass.
The situation has now changed completely,
following the development of the DD-HMC simulation 
algorithm [\ref{SchwarzI}--\ref{SchwarzIII}] 
and of a fine-tuned version of 
the Hasenbusch-accelerated HMC algorithm 
[\ref{Hasenbusch}--\ref{UrbachEtAl}], both being much faster
than the algorithms used thus far
(for related earlier studies of unquenched lattice QCD,
see refs.~[\ref{SesamBig}--\ref{OrthEtAl}], for example).
The success of these algorithms is partly
also due to the fact, discovered later [\ref{Stability}], 
that the Wilson--Dirac operator has a safe spectral gap
in the large-volume regime of QCD,
even though chiral symmetry is not 
exact in the Wilson theory.

In this paper we report on extensive simulations of two-flavour QCD
in a range of lattice spacings, lattice sizes 
and quark masses not explored before.
Apart from sect.~2, where the simulations are briefly described,
the emphasis is put on the main physics results.
Further details of the simulations, data tables, etc., 
will be published in a forthcoming more technical paper.

\section 2. Simulation table

As already indicated, the lattice theory is set up following Wilson
[\ref{Wilson}], with a doublet of sea quarks of equal mass.
In this theory, the sea quarks represent 
the up and down quarks, while
the strange quark will be added later, at the level of a valence quark,
i.e.~without the associated quark determinant.
 
Three series of lattices were simulated, labelled $A$,
$B$ and $D$ (see table~1)\kern2pt\footnote{$\dag$}{\footnotefont%
Our notation and normalization conventions coincide with those
of ref.~[\ref{OaImp}]. As usual we quote the values of 
$\beta=6/g_0^2$ and $\kappa=8+2m_0$ instead of the 
bare coupling $g_0$ and sea-quark mass $m_0$.}.
In the $D$ series, the Sheikholeslami--Wohlert 
term [\ref{SW}] was included in the quark action
with coefficient $\csw$ set to the value determined by 
the ALPHA collaboration [\ref{NPimp}], thus ensuring 
non-perturbative on-shell O($a$) improvement. 
The size of the representative
ensemble of gauge-field configurations generated in each case
is given in the last column of table~1. 

In physical units, the lattice spacing on the $A$, $B$ and
$D$ lattices is estimated to be $0.0717(15)$, $0.0521(7)$
and $0.0784(10)$~fm respectively (see sect.~4). 
The three series of simulations cover 
a similar range of sea-quark masses,
from values slightly above the strange-quark mass $\ms$
down to values close to $\ms/4$.

\topinsert
\newdimen\digitwidth
\setbox0=\hbox{\rm 0}
\digitwidth=\wd0
\catcode`@=\active
\def@{\kern\digitwidth}
\tablecaption{Simulations included in the physics analysis} 
\vskip1.5ex
$$\vbox{\settabs\+&%
                  xxxxxxx&xx&
                  xxxxxxxxx&xx&
                  xxxxx&xx&
                  xxxxxxxx&xx&
                  xxxxxxxx&x&
                  xxxxxxxxx&\cr
\thicktablerule
\vskip1.0ex
                \+& \hfill Run\hfill
                 && \hfill Lattice\hfill
                 && \hfill $\beta$\hfill
                 && \hfill $\csw$\hfill
                 && \hfill $\kappa$\hfill
                 && \hfill $\Ncfg$\hfill
                 &\cr
\vskip1.0ex
\thintablerule
\vskip1.2ex
  \+& \hfill $A_1$\hfill
  &&  \hfill $32\times24^3$\hfill
  &&  \hfill $5.6$\hfill 
  &&  \hfill $0$\hfill
  &&  \hfill $0.15750$\hfill
  &&  \hfill $@64$\hfill
  &\cr
\vskip0.3ex
  \+& \hfill $A_2$\hfill
  &&  \hfill $$\hfill
  &&  \hfill $$\hfill 
  &&  \hfill $$\hfill
  &&  \hfill $0.15800$\hfill
  &&  \hfill $109$\hfill
  &\cr
\vskip0.3ex
  \+& \hfill $A_3$\hfill
  &&  \hfill $$\hfill
  &&  \hfill $$\hfill 
  &&  \hfill $$\hfill
  &&  \hfill $0.15825$\hfill
  &&  \hfill $100$\hfill
  &\cr
\vskip0.3ex
  \+& \hfill $B_1$\hfill
  &&  \hfill $64\times32^3$\hfill
  &&  \hfill $5.8$\hfill 
  &&  \hfill $0$\hfill
  &&  \hfill $0.15410$\hfill
  &&  \hfill $100$\hfill
  &\cr
\vskip0.3ex
  \+& \hfill $B_2$\hfill
  &&  \hfill $$\hfill
  &&  \hfill $$\hfill 
  &&  \hfill $$\hfill
  &&  \hfill $0.15440$\hfill
  &&  \hfill $101$\hfill
  &\cr
\vskip0.3ex
  \+& \hfill $B_3$\hfill
  &&  \hfill $$\hfill
  &&  \hfill $$\hfill 
  &&  \hfill $$\hfill
  &&  \hfill $0.15455$\hfill
  &&  \hfill $104$\hfill
  &\cr
\vskip0.3ex
  \+& \hfill $B_4$\hfill
  &&  \hfill $$\hfill
  &&  \hfill $$\hfill 
  &&  \hfill $$\hfill
  &&  \hfill $0.15462$\hfill
  &&  \hfill $102$\hfill
  &\cr
\vskip0.3ex
  \+& \hfill $D_1$\hfill
  &&  \hfill $48\times24^3$\hfill
  &&  \hfill $5.3$\hfill 
  &&  \hfill $1.90952$\hfill
  &&  \hfill $0.13550$\hfill
  &&  \hfill $104$\hfill
  &\cr
\vskip0.3ex
  \+& \hfill $D_2$\hfill
  &&  \hfill $$\hfill
  &&  \hfill $$\hfill 
  &&  \hfill $$\hfill
  &&  \hfill $0.13590$\hfill
  &&  \hfill $171$\hfill
  &\cr
\vskip0.3ex
  \+& \hfill $D_3$\hfill
  &&  \hfill $$\hfill
  &&  \hfill $$\hfill 
  &&  \hfill $$\hfill
  &&  \hfill $0.13610$\hfill
  &&  \hfill $168$\hfill
  &\cr
\vskip0.3ex
  \+& \hfill $D_4$\hfill
  &&  \hfill $$\hfill
  &&  \hfill $$\hfill 
  &&  \hfill $$\hfill
  &&  \hfill $0.13620$\hfill
  &&  \hfill $168$\hfill
  &\cr
\vskip0.3ex
  \+& \hfill $D_5$\hfill
  &&  \hfill $$\hfill
  &&  \hfill $$\hfill 
  &&  \hfill $$\hfill
  &&  \hfill $0.13625$\hfill
  &&  \hfill $169$\hfill
  &\cr
\vskip1.2ex
\thicktablerule
}
$$
\vskip-0.0ex
\endinsert

All simulations were performed using
the DD-HMC algorithm [\ref{SchwarzIII}], which
combines domain decomposition ideas with 
the Hybrid-Monte-Carlo algorithm [\ref{HMC}] (hence the name DD-HMC).
As explained in refs.~[\ref{SchwarzIII},\ref{Dublin}], 
the speed of the algorithm also very much depends on
the use of the Sexton--Weingarten multiple-time integration scheme
[\ref{SextonWeingarten}].
In practice, a relevant performance figure is
the number $\Nops$ of floating-point operations
required for the generation of an ensemble of $100$ statistically
independent field configurations on a $2L\times L^3$ lattice
at a specified lattice spacing and sea-quark mass. A formula
that fits our experience with the DD-HMC algorithm well is
\equation{
   \quad\Nops=k
   \left({20\;\MeV\over\mbar}\right)
   \left({L\over3\;\fm}\right)^5
   \left({0.1\;\fm\over a}\right)^6\;\hbox{Tflops$\times$year},
   \enum
}
where $\mbar$ denotes the running sea-quark mass 
in the $\overline{\rm MS}$ scheme at renormalization scale $\mu=2$~GeV
and $k\simeq0.05$ if the O($a$)-improved theory is simulated
($k\simeq0.03$ without improvement). 

In 2001, at the annual conference on lattice field theory in Berlin, 
a similar formula was presented by Ukawa
[\ref{UkawaBerlin}], summarizing the experience made by the CP--PACS
and JLQCD collaborations with the 
algorithms available at the time.
With respect to that formula, the
scaling exponent of the quark mass in eq.~(2.1)
is reduced from $3$ to $1$, the exponent of the lattice spacing 
from $7$ to $6$, and the constant $k$ is roughly
$100$ times smaller.

Apart from the operations count,
the suitability of the simulation algorithm 
for parallel processing is a key issue, 
if large lattices are to be simulated. 
Domain de\-com\-position methods tend to perform well from this 
point of view, and one of the design goals of the 
DD-HMC algorithm was in fact 
to keep the communication overhead small 
[\ref{SchwarzII},\ref{SchwarzIII}].
Special-purpose
computers are then not required and most simulations 
(including the $B$ series)
were actually performed on commodity PC clusters
with up to $64$ double-processor nodes\kern1.5pt%
\footnote{$\dagger$}{\footnotefont%
The program that was used for the simulations of the 
O($a$)-improved theory (the $D$ series) can be downloaded from
{http://cern.ch/luscher/DD-HMC}.
}.

\section 3. Physical sea-quark effects

Once the sea quarks are included in the simulations, an obvious
question is whether their presence has a visible effect
on the computed quantities. 
Correlation functions of local fields 
depend on the sea-quark content of the theory
in various ways.
The flavoured pseudo-scalar densities,
for example, can only couple 
to multi-meson states through the creation of 
virtual quark pairs (see fig.~1).
Higher-states contributions to the correlation functions
of these fields
are therefore expected to increase 
when the sea-quark mass is 
lowered (thus moving away from quenched QCD).

\topinsert
\vbox{
\vskip0.0cm
\epsfxsize=8.0cm\hskip2.0cm\epsfbox{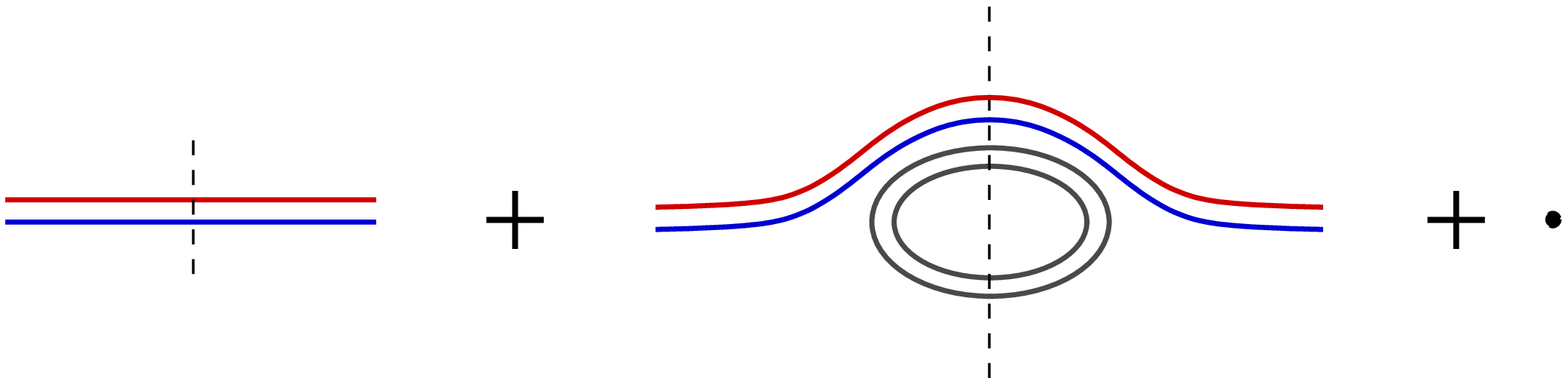}
\vskip0.3cm
\figurecaption{%
Quark-line diagrams contributing to the two-point correlation function (3.1).
In the second diagram, two pairs of sea-quarks are created from the vacuum
and bind into a pair of pseudo-scalar mesons. 
}
\vskip0.0cm
}
\endinsert

Indications for the presence of multi-meson intermediate states
in some two-point correlation functions were already found
some time ago by the UKQCD collaboration
[\ref{UKQCDlight}]. We now have data at smaller quark masses,
where the effect is more pronounced and where a significant
dependence on the sea-quark mass is seen
in both the pseudo-scalar and vector channels. 

For illustration we introduce two valence quarks, labelled $r$ and $s$,
and consider the two-point correlation function
\equation{
  \quad C(t)=-\int_{x_0=t}\rmd^3x
  \left\langle\left(\rbar\dirac{5}s\right)(x)
  \left(\sbar\dirac{5}r\right)(0)\right\rangle
  \enum
}
of the corresponding 
flavoured pseudo-scalar density at zero spatial momentum
(for simplicity we use a continuum notation in this section).
In finite volume, $C(t)$ may be expanded in a spectral series
\equation{
  \quad C(t)\mathrel{\mathop=_{t\to\infty}}
  c_0\rme^{-Mt}+c_1\rme^{-M't}+\ldots,
  \enum
}
where $M$ denotes the mass of the associated pseudo-scalar meson
and $M'$ the energy of a three-meson state with all particles at rest.

\topinsert
\vbox{
\vskip0.0cm
\epsfxsize=11.0cm\hskip0.5cm\epsfbox{plots/seaquark.eps}
\vskip0.3cm
\figurecaption{%
Simulation results (data points) for 
the effective pseudo-scalar mass $\Meff(t)$ in MeV as a function of
the time $t$ in fm (runs $D_2$ and $D_4$).
The average valence-quark masses $\hat{m}_{\rm val}$
and the meson masses $M$ (grey bands) are nearly the same
in the two cases, while the sea-quark mass $m_{\rm sea}$
changes by a factor of $2$ (quoted quark masses are bare current-quark masses).
}
\vskip0.0cm
}
\endinsert

Plots of the effective mass 
\equation{
  \quad \Meff(t)=-{\rmd \over\rmd t}\ln C(t)=M+c\,\rme^{-(M'-M)t}+\ldots
  \enum
}
now show that the higher-states contributions are not small in general.
Moreover, as is evident from fig.~2, they tend to grow when
the sea quarks become lighter. In the cases shown in the figure, the 
energy $M'$ of the lowest three-meson intermediate state
is expected to be approximately equal to $M+2\mpi$, where $\mpi$ 
denotes the mass of the pseudo-scalar mesons made of the sea quarks.
The two-state formula (3.3) actually fits the data quite well
if this expression is assumed and if $\mpi$ is determined
from the sea-quark pseudo-scalar correlation function 
(solid lines in fig.~2).

While the observed enhancement of higher-states contributions
is in line with qualitative theoretical expectations, their presence 
also tends to complicate the analysis of the simulation data.
In particular, at small sea-quark masses, the computation of hadron masses
may require accurate data at larger time separations than was the case 
in quenched QCD.
Multi-mass fits and variational methods can be helpful at this 
point, although the associated systematic uncertainties must then be 
balanced against the possibly lower statistical errors.

\section 4. Setting the scale

The choice of a physical reference scale
is an important step in the analysis of the simulation data.
Results obtained on different lattices can then be
expressed in units of this scale and thus be compared with
one another.

\topinsert
\vbox{
\vskip0.0cm
\epsfxsize=10.4cm\hskip0.5cm\epsfbox{plots/mref_pqcd.eps}
\vskip0.3cm
\figurecaption{%
On each lattice,
the masses of the $K$ and the $\Kstar$ were computed
at $4$ or $5$ values of the bare strange-quark mass.
The results obtained on the $D$ lattices are plotted here for
illustration (data points).
At the lighter sea-quark masses,
the point where $\MK/\MKstar=0.554$ (dotted line) is found 
by a quadratic interpolation in the strange-quark
mass of the nearest data points.
The solid line shows the interpolation 
in the case of the lattice $D_5$.
}
\vskip0.0cm
}
\endinsert

Following ref.~[\ref{OaImp}], we adopt a mass-independent 
scheme where the lattice spacing in physical units
is the same on all lattices at a given bare coupling.
Different choices of the reference scale are possible,
none of which appears to be free of some practical or conceptual 
shortcoming.
Here we add a valence strange quark to the theory and
determine the scale through
the pion mass and the masses of the pseudo-scalar and vector mesons 
that are made of a strange antiquark and a sea quark (we
refer to these as the $K$ and the $\Kstar$).
More precisely, 
we adjust the quark masses so that the ratios $\MK/\MKstar$ and
$\Mpi/\MK$ assume some prescribed values
and then take $\MK$ as the reference scale.

Since we wish to set the scale in a physically sensible way, we
require the ratio $\MK/\MKstar$ to be equal to its physical value
of $0.554$. This condition fixes the strange-quark mass $\ms$ 
at any given coupling and sea-quark mass $m$ (see fig.~3). 
Ideally we would like the latter to be such
that $\Mpi/\MK$ assumes its physical value too, but this would require
a long extrapolation in the sea-quark mass 
and, moreover, would be a point where the $\Kstar$ is unstable
(i.e.~the extrapolation would have to go through a kinematical threshold).

We now note, however, that once $\ms$ is fixed, 
the reference scale in lattice units, $a\MK$,
appears to be weakly dependent on $m$, particularly so at small $m$ 
(see fig.~3).
The reason for this behaviour
(which is seen on all series of lattices) could be that
both $\MK$ and $\MKstar$ are functions of 
$m+\ms$ rather than of $m$ and $\ms$ separately, up to 
corrections proportional to the squares of the masses.
In any case, the observation suggests the reference scale 
to be defined at
the point where, say, $\Mpi/\MK=0.85$, which is 
within the available data range.
This convention, although somewhat unphysical, 
is entirely satisfactory for the purpose
of comparing results from different lattices.

The results for the reference scale obtained in this way are
summarized in table~2. In order to avoid any confusions, we
mark all quantities evaluated at the reference point with 
a subscript ``ref''. The sea-quark and strange-quark hopping parameters
at the reference point, for example, are denoted by $\kappa_{\rm ref}$
and $\kappa_{s,{\rm ref}}$.
Setting $\MKref=495$ MeV,  
this leads to the lattice spacings quoted in the last column
of the table, while for the pion masses at the smallest sea-quark
masses on the 
$A$, $B$ and $D$ series of lattices we obtain $403$, $381$ and 
$377$ MeV respectively.

\topinsert
\newdimen\digitwidth
\setbox0=\hbox{\rm 0}
\digitwidth=\wd0
\catcode`@=\active
\def@{\kern\digitwidth}
\tablecaption{Determination of the lattice spacing\kern1pt$^\ast$} 
\vskip-1.5ex
$$\vbox{\settabs\+&%
                  xxxxxxxxxxxx&xx&
                  xxxxxxxxxxxxx&xx&
                  xxxxxxxxxxxxx&xx&
                  xxxxxxxxxx&xx&
                  xxxxxxxxxx&x\cr
\thicktablerule
\vskip1.0ex
                \+& \hfill Lattice series\hfill
                 && \hfill $\kappa_{\rm ref}@$\hfill
                 && \hfill $\kappa_{s,{\rm ref}}@$\hfill
                 && \hfill $a\MKref@@$\hfill
                 && \hfill $a$\kern1.5pt[fm]@\hfill
                 &\cr
\vskip1.0ex
\thintablerule
\vskip1.5ex
  \+& \hfill $A$\hfill
  &&  \hfill $0.15822(3)@@$\hfill
  &&  \hfill $0.15769(4)@@$\hfill 
  &&  \hfill $0.180(4)@@$\hfill
  &&  \hfill $0.0717(15)$\hfill
  &\cr
\vskip0.3ex
  \+& \hfill $B$\hfill
  &&  \hfill $0.154561(12)$\hfill
  &&  \hfill $0.154257(10)$\hfill 
  &&  \hfill $0.1310(17)$\hfill
  &&  \hfill $0.0521(7)@$\hfill
  &\cr
\vskip0.3ex
  \+& \hfill $D$\hfill
  &&  \hfill $0.136207(7)@$\hfill
  &&  \hfill $0.135912(13)$\hfill
  &&  \hfill $0.197(3)@@$\hfill
  &&  \hfill $0.0784(10)$\hfill
  &\cr
\vskip1.5ex
\thicktablerule
\vskip1.5ex
\+&{\footnotefont%
$^*$ At the quark masses where $\MK/\MKstar=0.554$ and $\Mpi/\MK=0.85$ 
}&\cr
}
$$
\vskip-2ex
\endinsert

The lattice spacings calculated here are 
significantly smaller than those previously published by us
in a conference report [\ref{Dublin}], where 
the Sommer radius [\ref{SommerRadius}] was used as reference scale.
Larger lattice spacings are also obtained 
if the scale is set by the $K$ and $\Kstar$ masses, similarly to
what was done here, but at larger sea-quark masses (see fig.~3).
As a result of the new
determination of the lattice spacings, our
estimates of the pion masses in MeV are pushed to
higher values than those quoted in ref.~[\ref{Dublin}].
Moreover, we decided to discard 
the simulation at the lightest quark mass reported there,
because the lattice turned out to be too small for that mass.

All this illustrates the fact that at present 
the assignment of physical units remains somewhat ambiguous. 
For a definitive solution of the problem,
simulations at smaller quark masses will probably be required, 
and the scale setting may eventually have to be
based on the properties of the stable hadrons 
(the pion and the nucleon in the two-flavour theory).

\section 5. Quark-mass dependence of $\mib\Mpi$ and $\mib\Fpi$

The bare quark masses that appear in the 
lattice action of the Wilson theory 
require a power-divergent additive renormalization.
This complication can be bypassed by extracting
the quark masses from the
PCAC relation (see ref.~[\ref{OaImp}], for example; in the 
improved theory,
we used the non-perturbatively 
improved axial current [\ref{NPimpCurrent}]).
Moreover, ratios of these masses do not need to be renormalized since
the multiplicative renormalization factor cancels.

In fig.~4 the ratio $(\Mpi/\MKref)^2$ 
is plotted as a function of the 
corresponding ratio of quark masses. If there were no systematic effects,
all data points shown in this figure would have to lie on a single curve,
within statistical errors,
representing the mass dependence of $(\Mpi/\MKref)^2$ in the continuum limit.
Note that the statistical errors of the points
are uncorrelated, except for the correlations that 
are introduced through the normalization factors. The quality of the 
empirical fit (solid line)
then suggests that no statistically significant systematic effects 
are seen in this plot.

\topinsert
\vbox{
\vskip0.0cm
\epsfysize=6.5cm\hskip0.1cm\epsfbox{plots/mpisq.eps}
\vskip0.3cm
\figurecaption{%
Dependence of the square of the pion mass $\Mpi$ on the sea-quark mass $m$.
The solid curve is a quadratic least-squares fit (with constant term)
of all data points, and the plot on the right is a blowup of the region
enclosed by the little box.
}
\vskip0.0cm
}
\endinsert

Another rather striking outcome is that $\Mpi^2$ is 
a nearly linear function of the sea-quark mass $m$ in
the range covered by the data. There is a visible curvature towards the larger 
masses in fig.~4, but the coefficient of the quadratic term in the empirical 
fit, $y=-0.03(3)+1.03(5)x+0.02(2)x^2$, 
is small. 
In the range $\Mpi/\MKref\leq1.1$, 
the data are also well represented by a straight line 
through the origin.

The corresponding plot of the pion decay constant $\Fpi$, given
in units of the decay constant $\FKref$ of the $K$ meson
at the reference point,
is more difficult to interpret (see fig.~5).
Apart from the fact that the 
statistical errors tend to be larger here, the results
of the $D$ series of simulations appear to be significantly different 
from those of the $A$ and $B$ series.
There is no obvious curvature in either set of data points, and 
correlated straight-line fits (solid lines) are found to be statistically
consistent. Although the two lines are visibly different,
the fitted values of their slopes, $0.235(11)$ and $0.192(11)$, 
deviate from each other by less than $3$ times the combined 
statistical error. The statistical significance of the
effect is thus not overwhelming.

When discussing systematic errors, an important
point to note is that the axial-current renormalization constant
$\ZA$ cancels in the ratio of decay constants plotted in fig.~5.
Moreover, the ratio is largely insensitive to the values of 
the improvement coefficients $\cA$ and $\bA$ 
[\ref{OaImp},\ref{NPimpCurrent}] on which the axial
current in the O($a$) improved theory depends.
Lattice effects may still be significant, however, and we can also not exclude  
the presence of important finite-volume effects. 
A more specific problem is that
any variations in the normalization factors 
$\FKref$ and $\mref+\msref$
propagate to the slopes of the lines in fig.~5.
The alignment of the data points may therefore appear to be 
better or worse,
depending on the statistical fluctuations at the reference point
and on its detailed specification.

\topinsert
\vbox{
\vskip0.0cm
\epsfysize=6.5cm\hskip0.5cm\epsfbox{plots/fpi.eps}
\vskip0.3cm
\figurecaption{%
Dependence of the pion decay constant $\Fpi$ on the sea-quark mass $m$.
The solid curves are linear least-squares fits of the data points from
the $A$ and $B$ lattices
(upper line) and of the points from the $D$ lattices (lower line).
}
\vskip0.0cm
}
\endinsert

The fact that the lines in fig.~5 have nearly the same
intercept in the chiral limit is probably an accident.
Both lines also practically 
pass through $\Fpi/\FKref=0.82$ (the experimental value
of $\Fpi/\FK$) at $\Mpi/\MKref=0.28$. However, as 
will be shown in the next section,
such extrapolations to smaller quark masses 
could be misleading.

\section 6. Comparison with chiral perturbation theory

In two-flavour QCD with unbroken isospin symmetry, 
the chiral expansion of the pion mass reads [\ref{GasserLeutwyler}]
\equation{
  \quad\mpi^2=M^2
  +{M^4\over32\pi^2 F^2}\ln(M^2/\Lambda_3^2)+\ldots,
  \qquad M^2\equiv2Bm,
  \enum
}
where $F$, $B$ and $\Lambda_3$ are a priori unknown constants.
A phenomenological analysis, taking low-energy experimental
data as input, suggests [\ref{GasserLeutwyler},\ref{ColangeloDurr}]
\equation{
  \quad F=86.2\pm0.5\,\MeV,
  \qquad
  \lbar_3\equiv
  \left.\ln(\Lambda_3^2/M^2)\right|_{M=139.6\,\MeV}=2.9\pm2.4,
  \enum
}
while $B$ depends on the renormalization scheme for 
the quark mass and thus cannot be determined from such data alone.
The chiral expansion of the pion decay constant
has the form [\ref{GasserLeutwyler}]
\equation{
  \quad\Fpi=F-{M^2\over16\pi^2 F}\ln(M^2/\Lambda_4^2)+\ldots\enum
}
and the phenomenological discussion leads to the estimate 
[\ref{ColangeloGasserLeutwyler}] 
\equation{
  \quad\left.\lbar_4\equiv\ln(\Lambda_4^2/M^2)\right|_{M=139.6\,\MeV}
  =4.4\pm0.2\enum
}
for the low-energy constant $\Lambda_4$.

In principle the low-energy constants 
can be determined from lattice data without recourse to 
phenomenological estimates. 
However, as will become clear shortly, 
the available simulation data are
insufficient for a solid analysis
of this kind. While the situation 
in the case of the pion mass is somewhat more favourable,
our principal goal in the following will be to find out 
whether the data are compatible with the expansions (6.1) and (6.3)
for a reasonable choice of the parameters.

We first need to rewrite the equations in a form where all dimensioned 
quantities are expressed in units of the scales at the reference point.
To this end, it is helpful to introduce the abbreviations
\equation{
   \quad
   x={2m\over\mref+\msref},
   \qquad 
   C={\MKref^2\over32\pi^2\FKref^2},
   \enum
}
and to define the scaled parameters
\equation{
  \quad\Fhat={F\over\FKref},
  \qquad 
  \Bhat={\mref+\msref\over \MKref^2}\,B,
  \qquad
  \lhat_n=\ln(\Lambda_n^2/\MKref^2).
  \enum
}
The chiral expansions 
\equation{
  \quad{\mpi^2\over\MKref^2}=\Bhat x
  +C{\Bhat^2 x^2\over \Fhat^2}\bigl\{\ln(\Bhat x)-\lhat_3\bigr\}
  +\ldots,
  \enum
  \next{2ex}
  \quad{\Fpi\over\FKref}=\Fhat
  -2C{\Bhat x\over \Fhat}\bigl\{\ln(\Bhat x)-\lhat_4\bigr\}
  +\ldots,
  \enum
}
may now be directly compared with the simulation data
(note that $\lhat_n=\lbar_n-2.53$ if $\MKref=495$ MeV is assumed).

\topinsert
\vbox{
\vskip0.0cm\noindent
\epsfysize=0.5\hsize\hskip1.7cm\epsfbox{plots/mpichpt.eps}%
\vskip0.3cm
\figurecaption{%
Fit of the quark-mass dependence of the square of 
the pion mass $\Mpi$
in the range $\Mpi/\MKref\leq1.1$, using the one-loop formula (6.7)
with $C=0.072$ and $\Fhat=0.70$
(solid line).
The fit is a correlated least-squares fit of all data points
in the specified range, 
with unconstrained parameters $\Bhat$ and $\lhat_3$.
}
\vskip0.0cm
}
\endinsert

The computation of the decay constant $\FKref$,
and thus of the constant $C$, involves
the renormalization constant $\ZA$ of the axial current. 
Recent estimates of the latter in the two-flavour Wilson theory
at the couplings of the $A$ and $B$ lattices are
$0.77(2)$ and $0.78(2)$ [\ref{BecirevicEtAl}], while in 
the case of the $D$ series of lattices we may use the value 
$0.75(1)$ determined by the ALPHA collaboration
[\ref{DellaMorteEtAl}].
For the constant $C$ we then find
$0.068(4)$, $0.071(4)$ and $0.076(3)$
respectively.
These figures are barely consistent with one another,
suggesting the presence of lattice or finite-volume
effects, but it should also be noted 
that the determination of $\ZA$ is not free of systematic ambiguities
[\ref{BecirevicEtAl},\ref{DellaMorteEtAl}]%
\kern2pt\footnote{$\dag$}{\footnotefont%
The same comments apply in the case of $\FKref$ 
where we obtain $107(3)$, $105(3)$ and $101(2)$ MeV
on the $A$, $B$ and $D$ lattices
(assuming $\MKref=495$ MeV as before).
The fact that these results are
all lower than the decay constant $\FK=113(1)$ MeV 
of the physical kaon may not be significant,
because the two-flavour theory neglects the 
effects of the strange sea quark.}.

\topinsert
\vbox{
\vskip0.0cm\noindent
\epsfysize=0.5\hsize\hskip2.2cm\epsfbox{plots/fpichpt.eps}
\vskip0.3cm
\figurecaption{%
Fit of the quark-mass dependence of the pion decay constant $\Fpi$
in the range $\Mpi/\MKref\leq1.1$, using
the one-loop formula (6.8) with $C=0.072$ and $\Bhat=1.106$
(solid line).
The dashed line with its $1$-sigma
error margin (grey band) represents an alternative fit 
that includes a hypothetical two-loop term.
}
\vskip0.0cm
}
\endinsert

A very accurate determination of $C$ is fortunately
not needed for the chiral fits, 
because $C$ only appears at next-to-leading order
in the chiral expansions.
We thus set $C=0.072$ and simplify the fit procedure
by substituting $\Fhat=0.70$ in eq.~(6.7),
which will turn out to be an approximately correct value.
In the range $\Mpi/\MKref\leq1.1$, 
the one-loop formula (6.7) then fits the data for the pion mass very well,
the fit parameters being
$\Bhat=1.11(6)(3)$ and $\lhat_3=0.5(5)(1)$ (see fig.~6;
the second errors are
estimates of the systematic uncertainty arising from the inaccurately
known values of $C$ and $\Fhat$). We did not attempt to estimate
the effects of any higher-order terms in eq.~(6.7) so that the 
quoted values of the fit parameters 
should be taken as effective values, describing the data in
the specified range of pion masses.

In the case of the pion decay constant,
the comparison of the simulation data 
with the chiral formula (6.8) is complicated by the 
scattering of the data points in fig.~5, which may partly 
be the result of systematic effects.
However, since the points line up at the smaller
quark masses, we may attempt to fit these,
setting $C$ and $\Bhat$ to the previously determined
values and adjusting $\Fhat$ and $\lhat_4$ 
(see fig.~7). 
The statistical quality of this fit (solid line) turns out to be quite good,
but the curvature of the fit function is not seen in the data and 
the fit therefore appears to be somewhat artificial.

A more plausible fit (dashed line) 
can be obtained
by including a hypothetical two-loop term proportional to
$\Bhat^2x^2/\Fhat^3$ in the chiral expansion (6.8), with a
coefficient $C'=0.046$ that is not unreasonably large.
The fit parameters $\Fhat$ and $\lhat_4$ change
from $0.60(4)$ and $1.6(1)$ to $0.73(3)$ and $0.73(8)$, respectively, 
when the two-loop term is added. 

The discussion in this section shows that
simulation data at significantly
smaller quark masses, with small systematic and statistical errors,
will be required for a reliable determination 
of the parameters in the chiral lagrangian.
It seems safe to conclude, however,
that our results in the range $\Mpi/\MKref\leq1.1$ are not
incompatible with chiral perturbation theory.
In particular, the fact that $\Mpi^2$ is a nearly linear
function of the quark mass $m$ is not in conflict with the 
presence of the chiral logarithm in eq.~(6.7).

\section 7. Concluding remarks

In the coming years, simulations of lattice QCD with Wilson quarks 
will no doubt rapidly progress towards
smaller quark masses and lattice spacings than are
reported here. In order to guarantee 
the stability of the simulations [\ref{Stability}],
but also to keep 
the finite-volume effects under control, the constraints
\equation{
  \Mpi L\geq3, \qquad L\geq 2\,\fm,
  \enum
}
should be respected in these computations. On a given
lattice, the bounds (7.1) set a lower limit on the 
lattice spacings and pion masses that can be reached (see fig.~8). 
Simulations of the two-flavour theory 
may not be practical at all these points,
but the cost formula (2.1) is encouraging and suggests that
simulations at $a\leq0.08$ fm and $\Mpi\leq300$ MeV, for example, 
can be performed already with the
computer resources available at present. However, to be able to sort
out the systematic errors, many lattices will have to be simulated
which may require a coordinated effort.

\topinsert
\vbox{
\vskip0.0cm
\epsfysize=0.44\hsize\hskip2.5cm\epsfbox{plots/limits.eps}
\vskip0.3cm
\figurecaption{%
Range of lattice spacings $a$ and pion masses $\Mpi$ defined by
the bounds (7.1) on a
$2L\times L^3$ lattice as a function of the lattice size $L/a$
(shaded area above the line labelled by the corresponding value of $L/a$).
}
\vskip0.0cm
}
\endinsert

On the $A$, $B$ and $D$ series of lattices, the smallest values of
$\Mpi L$ are
$3.5$, $3.2$ and $3.6$ respectively, while the spatial sizes $L$ of the 
lattices are estimated to be $1.72$, $1.67$ and $1.88$ fm, i.e.~somewhat
below the required minimum. Finite-volume effects
may not be totally negligible on these lattices 
and will need to be investigated, extending the
studies by Orth et al.~[\ref{OrthEtAl}]
to smaller quark masses and lattice spacings.
So far we did not include the nucleons 
in the physics analysis, because these are probably
even more sensitive to finite-volume effects
than the mesons.

It may be somewhat surprising that no significant
lattice effects were seen in fig.~4, even though
O($a$) counterterms were only included in the $D$ series 
of simulations. 
The weak dependence on the lattice spacing could be
related to the fact, first noted by
Sharpe and Singleton [\ref{SharpeSingleton}], that
the O($a$) lattice effects amount to
an additive quark-mass renormalization
at leading order of chiral perturbation theory.
Since the quark masses that appear in
the PCAC relation already
include all additive re\-nor\-ma\-li\-zations, it follows that
the data points plotted in fig.~4 are insensitive 
to these leading-order lattice effects. 

The mass dependence of the pseudo-scalar decay constant, on the other hand,
is a second-order effect in chiral perturbation theory.
At this order,
only some of the O($a$) lattice corrections 
can be compensated by a renormalization of the 
parameters in the chiral lagrangian,
and an accidental O($a$) improvement
is therefore not expected in this case.

\vskip1.0ex
We wish to thank Rainer Sommer for technical discussions
and Gilberto Colangelo, Stephan D\"urr, J\"urg Gasser and 
Heiri Leutwyler for some very helpful notes, summarizing 
some relevant results of chiral perturbation theory.
The numerical simulations were performed on PC
clusters at CERN, the Centro Enrico Fermi, the Institut f\"ur
Theoretische Physik der Universit\"at Bern (with a contribution from
the Schweizerischer Nationalfonds) and on a CRAY XT3 at the Swiss
National Super\-computing Centre (CSCS).
We are grateful to all these institutions for the continuous support
given to this project.

\beginbibliography


\bibitem{Wilson}
K. G. Wilson,
Phys. Rev. D10 (1974) 2445


\bibitem{SchwarzI}
M. L\"uscher,
JHEP 0305 (2003) 052

\bibitem{SchwarzII}
M. L\"uscher,
Comput. Phys. Commun. 156 (2004) 209

\bibitem{SchwarzIII}
M. L\"uscher,
Comput. Phys. Commun. 165 (2005) 199


\bibitem{Hasenbusch}
M. Hasenbusch,
Phys. Lett. B519 (2001) 177

\bibitem{HasenbuschJansen}
M. Hasenbusch, K. Jansen,
Nucl. Phys. B659 (2003) 299

\bibitem{DellaMorteEtAl}
M. Della Morte et al. (ALPHA collab.),
Comput. Phys. Commun. 156 (2003) 62

\bibitem{UrbachEtAl}
C. Urbach, K. Jansen, A. Shindler, U. Wenger,
Comput. Phys. Commun. 174 (2006) 87


\bibitem{SesamBig}
N. Eicker et al. (SESAM collab.),
Phys. Rev. D59 (1999) 014509



\bibitem{UKQCDbig}
C. R. Allton et al. (UKQCD collab.),
Phys. Rev. D60 (1999) 034507;
{\it ibid.} D65 (2002) 054502


\bibitem{CPPACSbig}
A. Ali Khan et al. (CP-PACS collab.),
Phys. Rev. Lett. 85 (2000) 4674 [E: {\it ibid.} 90 (2003) 029902];
Phys. Rev. D65 (2002) 054505 [E: {\it ibid.} D67 (2003) 059901]


\bibitem{JLQCDlight}
S. Aoki et al. (JLQCD collab.)
Phys. Rev. D68 (2003) 054502


\bibitem{CPPACSsmall}
Y. Namekawa et al. (CP-PACS collab.),
Phys. Rev. D70 (2004) 074503


\bibitem{OrthEtAl}
B. Orth, T. Lippert, K. Schilling,
Phys. Rev. D72 (2005) 014503


\bibitem{Stability}
L. Del Debbio, M. L\"uscher, L. Giusti, R. Petronzio, N. Tantalo,
JHEP 0602 (2006) 011


\bibitem{SW}
B. Sheikholeslami, R. Wohlert,
Nucl. Phys. B259 (1985) 572

\bibitem{OaImp}
M. L\"uscher, S. Sint, R. Sommer, P. Weisz,
Nucl. Phys. B478 (1996) 365


\bibitem{NPimp}
K. Jansen, R. Sommer (ALPHA collab.),
Nucl. Phys. B530 (1998) 185
[E: {\it ibid.} B643 (2002) 517]


\bibitem{HMC}
S. Duane, A. D. Kennedy, B. J. Pendleton, D. Roweth,
Phys. Lett. B195 (1987) 216


\bibitem{SextonWeingarten}
J. C. Sexton, D. H. Weingarten,
Nucl. Phys. B380 (1992) 665


\bibitem{Dublin}
M. L\"uscher, Lattice QCD with light Wilson quarks,
in: Proceedings of the 23rd International Symposium on Lattice Field Theory,
Dublin, 2005, eds. A. Irving, C. McNeile, C. Michael, PoS (LAT2005) 002


\bibitem{UkawaBerlin}
A. Ukawa, 
Nucl. Phys. B (Proc. Suppl.) 106--107 (2002) 195


\bibitem{UKQCDlight}
C. R. Allton et al. (UKQCD collab.),
Phys. Rev. D70 (2004) 014501


\bibitem{SommerRadius}
R. Sommer,
Nucl. Phys. B411 (1994) 839


\bibitem{NPimpCurrent}
M. Della Morte, R. Hoffmann, R. Sommer,
JHEP 0503 (2005) 029


\bibitem{GasserLeutwyler}
J. Gasser, H. Leutwyler,
Ann. Phys. 158 (1984) 142

\bibitem{ColangeloDurr}
G. Colangelo, S. D\"urr,
Eur. Phys. J. C 33 (2004) 543

\bibitem{ColangeloGasserLeutwyler}
G. Colangelo, J. Gasser, H. Leutwyler,
Nucl. Phys. B603 (2001) 125


\bibitem{BecirevicEtAl}
D. Be\'cirevi\'c et al. (SPQ$_{\rm CD}$R collab.),
Nucl. Phys. B734 (2006) 138

\bibitem{DellaMorteEtAl}
M. Della Morte et al. (ALPHA collab.),
JHEP 0507 (2005) 007


\bibitem{SharpeSingleton}
S. R. Sharpe, R. L. Singleton,
Phys. Rev. D58 (1998) 074501

\endbibliography

\bye